\documentclass[sort&compress
    ,final            
  ]
  {aipproc}

\layoutstyle{6x9}

\begin{document}

\title{Hadronization in nuclear DIS and ultra-relativistic HIC}

\author{T. Falter}{
  address={Physics Department, Brookhaven National Laboratory, 
Upton NY 11973, USA}
}
\author{W. Cassing}{
address={Institut fuer Theoretische Physik, Universitaet Giessen, 
D-35392 Giessen, Germany}
}
\author{K. Gallmeister}{
address={Institut fuer Theoretische Physik, Universitaet Giessen, 
D-35392 Giessen, Germany}
}
\author{U. Mosel}{
address={Institut fuer Theoretische Physik, Universitaet Giessen, 
D-35392 Giessen, Germany}
}

\begin{abstract}
We present a transport theoretical analysis of hadron attenuation in deep inelastic lepton scattering (DIS) off complex nuclei in the kinematic regime of the HERMES experiment. The HERMES data indicate the presence of strong prehadronic final state interactions shortly after the elementary lepton-nucleon interaction. The contribution of such (pre-)hadronic final state interactions to the observed jet quenching in ultra-relaticistic heavy ion collisions (HIC) at RHIC is estimated under the assumption that a similar space-time picture for hadronization also holds in a hot hadronic medium. Our results show that an additional mechanism for jet quenching must be at work for most central collisions.
\end{abstract}

\maketitle


Nuclear deep inelastic scattering (DIS) provides an excellent tool for the investigation of both the space-time picture of hadronization and prehadronic final state interactions (FSI) in a (cold) nuclear environment. Since the products of the initial $\gamma^*N$ interaction have to pass through the nucleus the nuclear target can be viewed as a kind of micro-detector that is located very close to the production vertex. Systematic nuclear DIS experiments might therefore clarify whether hadron attenuation in cold nuclear matter is primarily of partonic origin \cite{partonic} or due to the FSI of color-neutral prehadrons \cite{prehadronic}.

In Refs.~\cite{Shadowing} we have developed a method to describe high energy photon and lepton induced reactions in the framework of a transport model based on the Boltzmann-Uehling-Uhlenbeck (BUU) equation. In our approach we split the lepton-nucleus interaction into two parts: In step 1) the virtual exchange photon produces a final state $X$ which we determine using the Lund Monte Carlo generators PYTHIA and FRITIOF. In addition we account for nuclear effects like binding energies, Fermi motion, Pauli blocking and coherence length effects that lead to nuclear shadowing of high energy photons. In step 2) the state $X$ is propagated through the nuclear target using our BUU transport model \cite{Eff99,Fal04}. This allows for a probabilistic coupled-channel description which accounts for particle creation, annihilation and elastic scattering in the final state interactions.

In the Lund model the lepton-nucleon scattering leads to the excitation of hadronic strings that fragment into color neutral prehadrons due to the creation of quark-antiquark pairs from the vacuum. It then takes a formation time $t_f$ for a each prehadron to build up its hadronic wave function. In our default approach \cite{Fal04} we assume that the string fragments instantaneously into prehadrons and assume a universal formation proper time $\tau_f=0.5$ fm/c in the restframe of each prehadron. Due to time dilatation the formation time $t_f=\frac{E_h}{m_h}\tau_f$ in the rest frame of the nucleus strongly depends on the hadron energy $E_h$ and mass $m_h$. The prehadronic cross sections during the formation time are determined via a simple constituent quark model by rescaling the corresponding hadronic cross sections: $\sigma^*_\mathrm{prebaryon}=\frac{n_\mathrm{org}}{3}\sigma_\mathrm{baryon}$, $\sigma^*_\mathrm{premeson}=\frac{n_\mathrm{org}}{2}\sigma_\mathrm{meson}$. Here $n_\mathrm{org}$ denotes the number of "leading" quarks in the prehadron, i.e.~those that were originally contained in the (resolved) virtual photon or target nucleon. Consequently, the beam and target remnants can interact with the nuclear environment right after the primary $\gamma^*N$ interaction.

\begin{figure}		
  \includegraphics[height=.22\textheight]{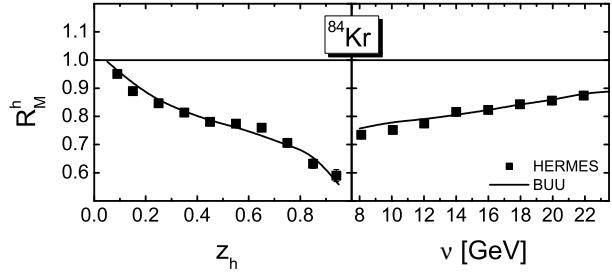}
  \caption{
  Multiplicity ratio $R_h^M$ (\ref{eq:multiplicity-ratio}) for charged hadron production
  on a $^{84}$Kr target in comparison with the HERMES data \cite{HERMES}.} \label{fig:fig1}
\end{figure}

The HERMES collaboration has extensively studied the attenuation of hadrons in scattering of a $E_\mathrm{beam}$=27.5 GeV positron beam on various gas targets \cite{HERMES}. The experimental observable is the multiplicity ratio:
\begin{equation}
	\label{eq:multiplicity-ratio}
R_M^h(z_h,\nu)=\frac{N_h(z_h,\nu)}{N_e(\nu)}\big|_A \bigg/ \frac{N_h(z_h,\nu)}{N_e(\nu)}\big|_D\,.
\end{equation}
Here $N_h$ denotes the yield of semi-inclusive hadrons in a given $(z_h,\nu)$-bin and $N_e$ the yield of inclusive deep inelastic scattering leptons in the same $\nu$-bin. The quantity $z_h=E_h/\nu$ is the photon energy fraction taken away by the hadron. In Fig.~\ref{fig:fig1} we show the multiplicity ratio (\ref{eq:multiplicity-ratio}) in a $^{84}$Kr target as function of the fractional energy $z_h$ and the photon energy $\nu$. Our model is in perfect agreement with the experimental findings. As we have shown in Ref.~\cite{Fal04} our model also successfully describes other aspects of the HERMES data also at beam energy $E_\mathrm{beam}$=12 GeV \cite{lowerEnergy}. The rescattering of color neutral prehadrons that are produced early after the initial $\gamma^*N$ interaction seems to be the predominant source for hadron attenuation in cold nuclear matter.

\begin{figure}
  \includegraphics[height=.22\textheight]{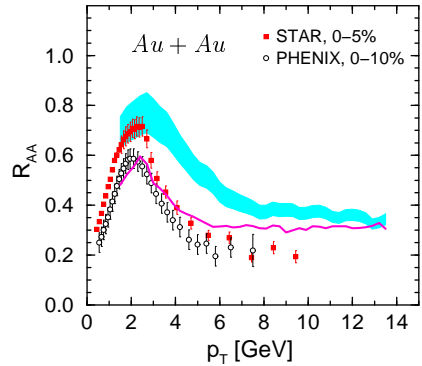}
  \caption{Comparison of the HSD calculation \cite{Gallmeister} 
  for the suppression factor $R_{\rm AA}$ (\ref{eq:ratioAA}) with the data for 
  charged hadrons from Ref.~\cite{PHENIXSTAR}. The Cronin
  effect has been accounted for in the calculation indicated by 
  the hatched band. The solid curve indicates the result without
  Cronin effect.}
\label{fig:fig2}
\end{figure}

In Refs.~\cite{Gallmeister} the contribution of prehadronic FSI to the observed jet quenching in ultra-relativistic heavy ion collisions at RHIC has been investigated in the framework of a similar transport model. The description of hadron formation and the prehadronic FSI in $AA$ collisions is the same as for the $eA$ reactions described above except for the additional assumption that prehadrons do not form at energy densities larger than 1 GeVfm$^{-3}$. The high $p_T$ (pre-)hadrons produced in a hard scattering event of a $AA$ collision have to traverse the hadronic gas produced in the soft interactions. The (pre-)hadronic rescattering then contributes to the experimentally measured nuclear suppression factor:
\begin{equation}
	  \label{eq:ratioAA}
  R_{\rm AA}(p_T) = \frac{1/N_{\rm AA}^{\rm event}\ d^2N_{\rm AA}/dy dp_T}
  {\left<N_{\rm coll}\right>/\sigma_{pp}^{\rm inelas}\ d^2
    \sigma_{pp}/dy dp_T}\ .
\end{equation}
While the model yields a nice description of jet quenching at mid-central collisions it clearly underestimates the jet suppression for most central collisions as can be seen from Fig.~\ref{fig:fig2}. This indicates the presence of an additional source of jet quenching in a possibly quark gluon plasma liquid created at most central heavy ion collisions at RHIC \cite{Andre}.

In our future studies of nuclear DIS we plan to implement a more realistic space-time picture of hadron production into our model \cite{Formation} and to investigate in greater detail hadron attenuation in the kinematic regime of the Jefferson ab experiment \cite{lowerEnergy,JLab}.


\end{document}